\documentclass[epj]{svjour}
\usepackage{amsmath,amssymb,amsfonts}
\usepackage{graphicx}
\usepackage{color}

\newcommand{\be}{\begin{equation}}
\newcommand{\ee}{\end{equation}}
\newcommand{\bea}{\begin{eqnarray}}
\newcommand{\eea}{\end{eqnarray}}
\newcommand{\bm}{\bibitem}
\newcommand{\bfg}{\begin{figure}}
\newcommand{\efg}{\end{figure}}
\newcommand{\bt}{\boldsymbol{\tau}}
\newcommand{\bp}{\boldsymbol{\phi}}
\newcommand{\al}{\alpha}
\newcommand{\gm}{\gamma}

\newcommand{\ep}{\epsilon}
\newcommand{\de}{\delta}

\newcommand{\om}{\omega}

\newcommand{\Th}{\Theta}
\newcommand{\lm}{\lambda}

\newcommand{\sg}{\sigma}
\newcommand{\Sg}{\Sigma}
\newcommand{\gmu}{\gamma^\mu}
\newcommand{\gmd}{\gamma_\mu}
\newcommand{\gz}{\gm^0}
\newcommand{\gf}{\gamma_5}
\newcommand{\us}{u \!\!\! /}
\newcommand{\ps}{p \!\!\! /}
\newcommand{\vp}{\vec p}
\newcommand{\vq}{\vec q}
\newcommand{\vx}{\vec x}
\newcommand{\dm}{\partial^\mu}

\newcommand{\etb}{\bar\eta}

\newcommand{\cz}{{\cal Z}}
\newcommand{\la}{\langle}
\newcommand{\ra}{\rangle}
\newcommand{\rw}{\rightarrow}
\newcommand{\F}{F_\pi}
\newcommand{\mn}{\mu\nu}

\newcommand{\on}{\overline{n}}
\newcommand{\osi}{\overline{\psi}}
\newcommand{\opi}{\overline{\Pi}}
\newcommand{\osg}{\overline{\sg}}
\newcommand{\wpi}{\widetilde{\Pi}}
\begin{document}

\title{QCD sum rule for nucleon in nuclear matter}

\author{S. Mallik\inst{1}\fnmsep\thanks{\email{mallik@theory.saha.ernet.in}}
\and Sourav Sarkar\inst{2}\fnmsep\thanks{\email{sourav@veccal.ernet.in}}}
\institute{Theory Division, Saha Institute of Nuclear Physics, 1/AF 
Bidhannagar, Kolkata 700064, India
\and Variable Energy Cyclotron Centre, 1/AF, Bidhannagar, 
Kolkata, 700064, India} 
 
\date{\today}

\abstract{We consider the two-point function of nucleon current in nuclear
matter and write a $QCD$ sum rule to analyse the residue of the nucleon pole 
as a function of nuclear density. The nucleon self-energy needed for the sum 
rule is taken as input from calculations using phenomenological $N\!N$ potential. 
Our result shows a decrease in the residue with increasing nuclear density,
as is known to be the case with similar quantities.}


\maketitle

\section{Introduction}

An important topic in strong interaction at non-zero temperature and
chemical potential is the propagation of hadrons through these media. Their
modified couplings and self-energies are useful not only in analysing the
experimental data on heavy-ion collisions, but also in extracting
indications of an eventual phase transition in the medium.

At low density, such a propagation in the hadronic phase can, in principle,    
be studied by invoking chiral perturbation theory to evaluate the appropriate  
two-point function \cite{Gasser1,Gasser2}. This effective theory of $QCD$ is of 
particular advantage at low temperature (and zero chemical potential), when the 
heat bath is dominated by pions. Their couplings with themselves and with other 
hadrons are highly constrained by the chiral symmetry of (massless) $QCD$. Thus 
not only are the vertices with pions related to those without the pions, but also 
they are suppressed by powers of pion momenta. Using these vertices, one can get 
the hadron parameters to a good accuracy by evaluating only a few relevant Feynman 
graphs. It has been applied with much success in calculating the properties of the 
pion and the nucleon at finite temperature
\cite{Gasser3,Gerber,Goity,Leutwyler1,Schenk,Toublan}.

In this work we are interested in finding the effect of strong interaction
on the propagation of nucleon at non-zero nucleon chemical potential ($\mu$)
i.e. in nuclear matter, at zero temperature ($\beta^{-1}=0$). Quite generally,
this propagation is studied by considering the two-point function,
\be
\Pi(E,\vec p)=i\int dtd^3x \,e^{i(Et-\vp\cdot\vx)}\la T\eta(x)\etb(0)\ra
\ee 
of a three-quark current $\eta(x)$, having the
quantum numbers of the nucleon \cite{Ioffe1,Chung}. Here $\la \cdots\ra$ denotes 
ensemble average: For any operator $O$, 
\[ \la O \ra =Tr[e^{-\beta(H-\mu N)}O]/\cz,~~~~~~\cz=Tr\,e^{-\beta(H-\mu
N)}\]
where $H$ and $N$ are the Hamiltonian and the number operator of the system.

Unfortunately, a straightforward calculation of Feynman graphs in this case, 
similar to the one at finite temperature, is not possible at present. The 
difficulty is due to the appearance of new and presumably large couplings. They 
are shown in Fig.~1, where the complete $\eta$-nucleon vertex and the
complete nucleon self-energy are analysed in terms of low order perturbative
vertices, whose structures may be obtained by the methods of chiral
perturbation theory. 

Thus, up to terms proportional to a single nucleon field $\psi (x)$, the 
current $\eta (x)$ is given by \cite{Mallik1} 
\be
\eta (x)= \lm\left( 1+\frac{i\bp(x)\cdot\bt}{2\F}\gf+\cdots\right)
\psi (x)\,,
\ee
where $\bp (x)$ is the pion field and the dots stand for terms with
increasing number of pion fields. Here $\F$, the so-called pion decay constant, 
is defined by the vacuum-to-pion matrix element of the axial-vector current 
$A_\mu^i (x)$,
\be
\la 0|A_\mu^i (x)|\pi^j (k)\ra =i\de^{ij}k_\mu\F e^{ik\cdot
x}\,,~~~\F=93\,\mathrm{MeV}\,,
\ee
just as $\lm$ is defined by the vacuum-to-nucleon matrix element of the
nucleon current $\eta (x)$
\be
\la 0|\eta (x)|N(p)\ra = \lm u(p) e^{ip\cdot x},
\ee
where $u(p)$ is a positive energy Dirac spinor. The value of $\lm$ is
obtained from $QCD$ sum rules for nucleon in vacuum \cite{Ioffe1}, 
\be
\lm^2=(1.2\pm0.6)\times 10^{-3} \mathrm{GeV}^6\,.
\ee
Terms in $\eta$ proportional to $\osi\psi\psi$ may also be obtained in the same way, 
bringing in two more new coupling constants \cite{Mallik1}. We now see in Fig.~1 
that unlike the vertex $\eta\osi\phi$ in graph (b), which is related to the vertex 
$\eta\osi$ of graph (a) itself, the vertex $\eta\osi\psi\osi$ in graph (c) is 
unknown and unlikely to be small. 

\bfg
\includegraphics[scale=0.9]{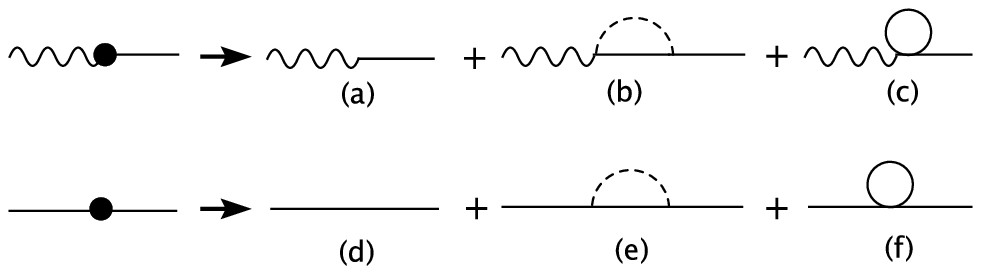}
\caption{Complete vertex and complete self-energy (denoted by solid circles)
analysed in terms of loop graphs.}
\efg

For the self-energy graphs, we note the pion-nucleon interaction Lagrangian 
\cite{Becher},
\be
{\cal{L}}_{int}=-\frac{g_A}{2\F}\osi(x)\gmd\gf\tau^i\psi(x)\dm\phi^i(x)
+\cdots,
\ee
where $g_A$ is the axial-vector coupling constant of the nucleon, $g_A=1.26$.
We again see in Fig.~1 that unlike the self-energy graph (e), which can be 
calculated with this Lagrangian, the other graph (f) poses difficulty, even though
chiral symmetry dictates the form of the four-nucleon effective Lagrangian,
whose coefficients can be fixed by the $N\!N$ scattering lengths
\cite{Weinberg}. Indeed, if one does calculate the graph (f) with this
four-nucleon interaction, one gets an unacceptably large value for the
nucleon self-energy \cite{Politzer}. The problem can be traced to the fact
that there are bound and virtual states very close to threshold in the
$N\!N$ system.  

In view of these difficulties, we give up calculating the nucleon
self-energy and content ourselves with evaluating only the nucleon pole residue,
by writing a $QCD$ sum rule for an appropriate amplitude representing the 
two-point function. (There are several approaches to $QCD$ sum rules in
medium \cite{Pasupathy,Drukarev,Cohen}. The one closest to ours is of
Ref.\cite{Cohen}.) To determine the nucleon pole term in the medium, we
still need the nucleon self-energy, which we take from the variational 
and the Brueckner type calculations using the phenomenological $N\!N$
interaction potentials \cite{Brockmann,Haar}. The contributions from the
remaining low energy singularities (branch cuts) are obtained by evaluating
the relevant Feynman graphs. 

Such sum rules in medium are usually written for complete amplitudes, including
their contributions from the vacuum. As noted in Ref.\cite{Jin}, a sum rule of 
this type cannot, however, determine the pole residue quantitatively.
The reason is that in such sum rules the vacuum contributions dominate over the
density dependent ones, at least at low densities, making the latter appear as 
non-leading terms.

In this work we {\it subtract} out the vacuum sum rule from the corresponding
one in medium, that is, we exclude the vacuum contributions from both the   
spectral and the operator sides. While this exclusion of the leading terms  
is expected to make the sum rule more sensitive to the medium dependent     
quantities, it also calls for a more careful scheme to saturate it.

We construct the spectral and the operator sides of the sum rule in Sec. 2
and 3 respectively and write the sum rule for the nucleon pole residue in Sec. 4. 
It is evaluated in Sec. 5. Our discussions are contained in Sec. 6.

\section{Spectral side of sum rule}

The low energy part of the spectral side of the sum rule is constructed
from singularities generated by the low mass physical particles. Besides the
complete nucleon pole, we also include branch cuts from $\pi N $ exchange.
We are thus led to consider the contributions of graphs of Fig.~2. As already
stated, the solid circles in graphs (a), (b) and (c) are not for diagrammatic
evaluation: The one representing the current-nucleon coupling is left as
unknown, to be determined by the sum rule. The other standing for the
self-energy is taken from the results in Ref.\cite{Brockmann,Haar}. The
remaining graphs do not involve any new couplings. Concerning graphs (e) and (f) 
we take only their non-pole parts, as the pole part is 
already included in graphs (a) and (b).

We shall calculate amplitudes in the real time version of the field theory in medium, 
where a two-point function assumes the form of a $2\times 2$ matrix \cite{Niemi}. 
(This formulation is reviewed in \cite{Landsmann}). But the dynamics is given 
essentially by a single analytic function, obtained by diagonalising the original matrix. 
Thus if $\Pi_{11}(E,\vp)$ is the $11$-component of the original matrix amplitude, the 
corresponding analytic function, to be denoted by the same symbol as in Eq.~(1), has the 
spectral representation \cite{MS1},
\be
\Pi(E,\vp)=\int_{-\infty}^{\infty}\frac{dE'}{2\pi}
\frac{\sg(E',\vp)}{E'-E-i\eta\ep(E')}
\ee
where the spectral function is related to the imaginary part of $\Pi_{11}$ by
\be
\sg(E,\vp)=2 \mathrm{coth}\{\beta(E-\mu)/2\}\, \mathrm{Im}\Pi_{11}(E,\vp).
\ee
For generality we calculate the amplitudes retaining both $\mu$ and $\beta$
and take the limit of zero temperature later.

\bfg
\includegraphics[scale=0.75]{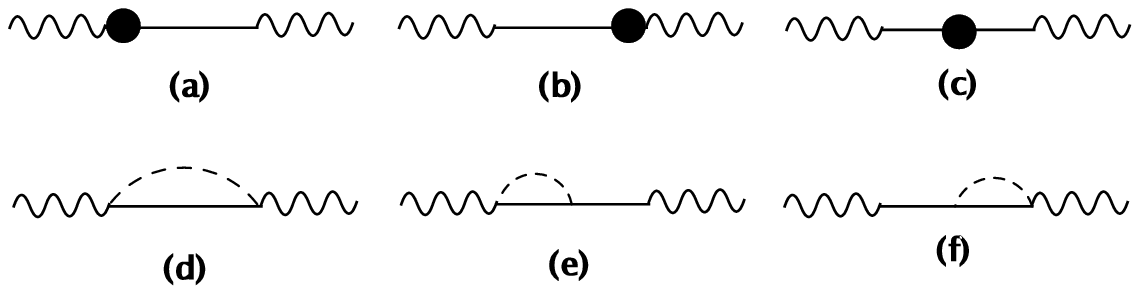}
\caption{Graphs representing the two-point function in the low
energy region.}
\efg

The two-point function due to the free propagation of nucleon, namely 
\[-\,\frac{\lm^2}{\ps-m+i\ep}\]
is modified by the vertex and the self-energy corrections of Figs.~2 (a),
(b) and (c) to
\be
\Pi (E,\vp)|_{(a+b+c)}=-\,\frac{\lm^{*2}}{\ps-m-\Sg(p)}
\label{effprop}
\ee 
where $\lm^*$ is the modified coupling parameter in nuclear matter and 
$\Sg(p)$ is the nucleon self-energy acquired in this medium.
 
In this work we restrict to $\vec p=0$, when $\Sg$ has the simple Dirac decomposition,
$\Sg=\Sg_S+\gz\Sg_V$. Then Eq.~(9) may be rewritten as
\be
\Pi(E)|_{(a+b+c)}=-\lm^{*2}\frac{\gz(E-\Sg_V)+m^*}{(E-m_1)(E-m_2)}\,.
\ee 
The scalar part $\Sg_S$ of the self-energy changes the mass $m$ of the free 
particle to the effective mass $m^*$ in the medium, $m^*=m+\Sg_S$. Similarly the 
vector part $\Sg_V$ shifts the rest energies, $\pm m$ of the free nucleon and the 
antinucleon, to $m_1=m^*+\Sg_V$ and $m_2=-m^*+\Sg_V$ respectively of the corresponding 
quasi-particles. Following our discussion in the Introduction, we work with the 
{\it subtracted} nucleon pole term,
\be
\opi(E)|_{(a+b+c)}=\Pi(E)|_{(a+b+c)}+\lm^2\frac{\gz E+m}{E^2-m^2}
\label{nucpole}
\ee
Here and below we use a {\it bar} over amplitudes $\Pi$ and spectral densities $\sg$  
to denote subtraction of the corresponding vacuum contributions. 

We now evaluate the remaining graphs of Fig.~(2). The imaginary part of the $11$
component of the matrix amplitude for graph (d) is given by
\cite{MS1,Weldon,Kobes}
\bea
&&\frac{{\mathrm{Im\,\Pi}}(E)_{11}|_{(d)}}{{\mathrm{tanh}}[\beta (E-\mu)/2]}
=-\left(\frac{3\lm^2\pi}{4F_\pi^2}\right)
\int\frac{d^3q}{(2\pi)^3 4\om_1\om_2}\times\nonumber\\
&&[(-\gz\om_1+m)\{(1-n_++n)\de(E-\om_1-\om_2)\nonumber\\
&&+(n_++n)\de(E-\om_1+\om_2)\}
-(\om_{1,2}\rw -\om_{1,2},\, n_+\rw n_-)]\nonumber\\
\eea
where $\om_1=\sqrt{m^2+\vq^{\,2}}$,~~~$\om_2=\sqrt{m_\pi^2+\vq^{\,2}}$ and $n_\pm$ and
$n$ are respectively the distribution functions for nucleons, antinucleons and
pions,
\be
n_\pm(\om_1)=\frac{1}{e^{\beta(\om_1\mp\mu)}+1}~,~~~~~~
n(\om_2)=\frac{1}{e^{\beta\om_2}-1}
\ee
Terms without the $n$'s are the vacuum contributions, which we subtract out, as
already stated above. Further, we restrict to zero temperature so that we have
to calculate only the term proportional to $n_+\to\theta(\mu-\om_1)$: we thus
get
\be
\osg(E)|_{(d)}=\frac{3\lm^2}{16\pi\F^2E}\sqrt{\om^2-m^2}(\gz\om-m)
\ee
on the Landau cut and the negative of the same quantity on the unitary cut.
Here $\om$ is the nucleon energy expressed in terms of the total energy $E$,
$\om=(E^2+m^2-m_\pi^2)/(2E)$. Because of the $\theta$-function in the integrand, 
the two cuts originally over the regions $0 \le E \le m-m_\pi$ (Landau) and
$ m+m_\pi \le E \le \infty$ (unitary) shrink respectively to
\be 
\mu-\sqrt{\mu^2-m^2+m_\pi^2}\le E\le m-m_\pi
\ee 
and
\be 
m+m_\pi\le E\le \mu+\sqrt{\mu^2-m^2+m_\pi^2}. 
\ee
Then the desired analytic function in the form of Eq.~(7) is given by
\be
\opi(E)|_{(d)}=\int_C\frac{dE'}{2\pi}\frac{\osg (E')|_{(d)}}{E'-E-i\eta\ep(E')}
\ee
where the subscript $C$ denotes the difference of two integrals,
\be
\int_C=\int_{\mathrm{Landau}}-\int_{\mathrm{unitary}}
\ee
Similarly the graphs (e) and (f) give 
\be
\opi(E)|_{(e+f)}=\frac{2(\gz E+m)}{E^2-m^2+i\ep} \int_C\frac{dE'}{2\pi}
\frac{\osg (E')|_{(e+f)}}{E'-E-i\eta\ep(E')}
\ee
where
\be
\osg (E)|_{(e+f)}=\frac{3\lm^2 g_A}{16\pi\F^2E}\sqrt{\om^2-m^2}
\{(m^2-E\om)+\gz m(E-\om)\}
\ee
Of course, we have yet to subtract out from Eq.~(19) its nucleon pole part, as 
the complete nucleon pole contribution is already represented by Eq.~(\ref{nucpole}).

At this point we adopt an improved choice of the amplitude following the suggestion 
in Ref. \cite{Jin}. Of the two quasi-particle poles, the one corresponding to the 
antinucleon cannot be treated as having narrow width, as we do in Eqs.~(10) and
(11)), because it can be annihilated strongly in nuclear medium. This 
ill-represented contribution may be suppressed by working with a modified amplitude.  
We split $\opi$ into even and odd parts,
\be
\opi(E)=\opi^{(e)}(E^2)+E\opi^{(o)}(E^2),
\ee
and deal with the combination,
\be
\wpi (E^2)=\opi^{(e)}(E^2)-m_2\opi^{(o)}(E^2)
\ee
(Recall that $m_2$ is the pole position of the quasi anti-nucleon.)
Indeed, if we separate the unsubtracted pole amplitude $\Pi(E)$ given by Eq.~(10) 
into even and odd parts and combine them as in Eq.~(22), we get the amplitude
\[-\frac{2\lm^{*2}m^*}{E^2-m_1^2}\frac{1}{2}(1+\gz)\]
where the pole at $E=m_2$ is removed to $E=m_1$. Note also that it is
proportional to $\frac{1}{2}(1+\gz)$. As we are interested in  a sum rule
for $\lm^*$, we shall project all amplitudes on to this combination in Dirac
space \footnote{The amplitudes proportional to $\frac{1}{2}(1-\gz)$ are             
expected to be relatively small and so comparable to those left out in our
approximation. Thus the corresponding sum rule may not be reliable, besides the 
fact that it will not involve $\lm^*$.}

We thus get the amplitudes corresponding to the different graphs of Fig.~(2) as
\bea
&&\wpi (E^2)|_{(a+b+c)}=-\frac{2\lm^{*2}m^*}{E^2-m_1^2}
+\frac{\lm^2(m-m_2)}{E^2-m^2}\\
&&\wpi (E^2)|_{(d)}=\frac{3\lm^2}{32\pi^2\F^2}\int_C
\frac{dE'}{E'}\frac{f(E')}{E'^2-E^2}\\
&&\wpi (E^2)|_{(e+f)}=-\frac{3\lm^2g_A}{16\pi^2\F^2} \int_C
\frac{dE'}{E'}\frac{E'+m}{E'-m}\frac{f(E')}{E'^2-E^2}~.\nonumber\\
\eea
Here $f(E)$ is given by
\be
f(E)=(E-m_2)\sqrt{\om^2-m^2} (\om -m)
\ee
In the last amplitude we have removed the nucleon pole contribution from graphs (e) 
and (f). The sum of these amplitudes will give the low energy part of the
spectral side of the sum rule.

\section{Operator side of sum rule}

In obtaining the operator product expansion, 
we need the explicit form of the quark current $\eta(x)_{D,i}$ with spin and 
isospin indices $D$ and $i$. Of the two independent possibilities involving three 
quark fields, we choose here the one, which for proton $(i=1)$ is
\footnote{Let us discuss here the preference, if any, for this choice. The
current written above corresponds to an axial-vector diquark coupled to a quark. 
Similarly, the other possibility, where $\gmd$'s are replaced by $\sg_{\mn}$'s, 
corresponds to an axial-tensor diquark coupled to a quark. Reliability of
results from $QCD$ sum rules requires that the nucleon pole term dominate
the continuum contribution and at the same time the higher order power
corrections be small compared to the leading ones retained in the sum rule.
On the basis of these criteria, one may choose the current with $\gmd$'s \cite{Ioffe2}.
There is a different line of investigation of the three-quark currents based
on models with instanton dominated (non-perturbative) vacuum. For a review
see Ref. \cite{Shuryak}. Each of the two currents mentioned above may be
written as a linear combination of two other three-quark currents, with
{\it relative coefficients} $\pm 1$, where they couple a scalar and a
pseudo-scalar diquark with a quark. These models find a large coupling of the
nucleon with the current containing the scalar diquark compared to the one
containing the pseudoscalar diquark. Thus they tend to support the use of any
of the original forms (with $\gmd$ and $\sg_{\mn}$) in writing the $QCD$
sum rules.},  
\[\eta(x)_{D,1}=\ep^{abc}(u^{aT}(x)C\gmu u^b(x))(\gf\gmd d^c(x))_D\,,\]
where $C$ is the charge conjugation matrix and $a,b,c$ are the colour indices. 
Because of subtraction of the vacuum amplitude, the unit operator in the 
operator product does not appear in $\opi (E)$. As is well-known, at higher 
dimensions, there are two sets of contributing operators in the in-medium sum rule: 
the old set, appearing already to the vacuum sum sum rule and the new set, involving
$u^{\mu}$, the four-velocity of the medium. From the ensemble average of the 
old set of operators, we have to subtract out the contribution of the vacuum
state.
 
We denote by $q$ any of the $u$ and $d$ quark flavours. Then the contributing 
operators of lowest dimension are $\overline{q} q$, $\overline{q}\us q$ 
(=$q^\dagger q$ in the rest frame of matter, where $u^\mu=(1,\vec 0)$). 
Next, there are the operators of dimension four, namely
$G^2=(\alpha_s/\pi)G_{\mu\nu}^aG^{\mu\nu a}$ and $\Theta^{f,g}\equiv u^\mu
u^\nu\Theta^{f,g}_{\mu\nu}$, where the (traceless) energy-momentum tensors of quarks 
and gluons are given respectively by
\bea
&&\Theta^f_{\mu\nu}=\sum_{q=u,d}\left(i\overline{q}\gm_\mu D_\nu q
-\frac{\hat{m}}{4}g_{\mu\nu}\overline{q} q\right),\\
&&\Theta^g_{\mu\nu}=-G^c_{\mu\lm}G^{\lm
c}_\nu+\frac{1}{4}g_{\mu\nu}G_{\alpha\beta}^cG^{\alpha\beta c}
\eea
Here $\hat{m}$ is the average quark mass of $u$ and $d$ quarks and
$G_{\mu\nu}^a$ are the gluon field strengths. Of the remaining operators we 
retain only the four-quark operators. Below we shall work in the limit of 
SU(2) flavour symmetry; thus $\la 0|\bar u u|0\ra = \la 0|\bar d d|0\ra\,,~~
\la\bar u u\ra = \la \bar d d\ra$ , etc.
 
In terms of the above operators, the operator expansion gives
\bea
\Pi(E,\vec 0)&&\stackrel{OPE}{\longrightarrow}  \frac{1}{4\pi^2}(\la\bar u u\ra 
+4\la  u^\dagger u\ra\gz)E^2\ln(-E^2/\mu^2)\nonumber\\
&&-\frac{1}{6\pi^2}\left( \frac{3}{16}\la G^2\ra +5\la\Theta^f\ra\right)\gz 
E\ln(-E^2/\mu^2)\nonumber\\
&& -\frac{2E}{3E^2}(\gz \la\bar u u\ra^2+2 \la\bar u u\ra \la u^\dagger u\ra)
\eea
with coefficients to zeroth order in $\al_s$ \footnote{The lowest order
coefficient of $\Th^g$ is proportional to $\al_s \ln(-E^2/\mu^2)$}. 
The renormalization scale $\mu$ is taken at $1$ GeV.

The four-quark condensates in medium encountered above have been factorised in 
a manner similar to that by saturating the vacuum condensates with the vacuum
intermediate state -- an approximation that may not be as good as in vacuum.
To rectify the error, we adopt the suggestion of Ref. \cite{Jin} to interpolate 
the factorised $\la \bar u u\ra^2$ between its values in vacuum and in medium.
Thus we shall replace it as
\be
\la \bar u u\ra^2 \to (1-f)\la 0|\bar u u|0\ra^2+f \la\bar u u\ra^2
\ee
where $f$ is a real parameter in the range $0 \leq f \leq 1$.
 
Recalling that the combination (22) of amplitudes is a function of $E^2$ ,
we can go to the spacelike region by setting $E^2=-Q^2,\,  Q^2>0$. 
While $\mu^2=1$ GeV$^2$ is a natural scale for the expectation values of the
operators, it is not convenient for the coefficients, as they contain powers 
of $\ln(Q^2/\mu^2)$ in higher orders. As is well-known in the context of
deep inelastic scattering, it is possible to use the renormalization group
equation to get rid of these (large) logarithms by shifting the scale for
the coefficients from $\mu^2$ to $Q^2$. The process brings in the (small)
anomalous dimensions of the operators; also the operators $\Th^q$ and
$\Th^g$ will mix \cite{Gross,Mallik2}. However, as we shall vary the Borel mass 
in the neighbourhood of $1$ GeV, the renormalization group improvement will give
small corrections, which we shall ignore in the present analysis.  

The nucleon number density $\overline{n}$ is related to the Fermi momentum
$p_F$ by ${\overline n}=2p_F^3/(3\pi^2)$. In normal nuclear matter, it 
is given by ${\overline n}_0=(110\, \mathrm{MeV})^3$ corresponding to 
$p_F=270\, \mathrm{MeV}$. To first order in $\overline n$, the change in the 
expectation value of an operator $O$ in nuclear matter relative to that in vacuum 
is given by its nucleon matrix element as
\be
\la O \ra =\la 0|O|0\ra +\frac{\la p|O|p\ra}{2m}\on
\ee
We apply this equation to the different operators. For $\bar u u$
and $u^\dagger u$, we get
\be
\la \bar u u\ra =\la 0|\bar u u|0\ra + \frac{\sg}{2\hat m}\on,~~~~
\la u^\dagger u\ra =\frac{3}{2}\on
\ee
where $\sg$ is the so-called nucleon $\sg$-term, \cite{Sainio},
\be
\sigma=\hat m\la p|\bar u u|p\ra/m=45\pm 8 \,\mathrm{MeV}\,.
\ee
The quark mass and the vacuum condensate are related by the Gell-Mann,
Oakes and Renner formula \cite{Gell-Mann},
\be
\F^2 m_\pi^2=-2\hat{m} \la 0|\bar{u} u|0\ra \,.
\ee
Two determinations of these quantities exist in the literature, namely
\bea 
&& \hat{m}=7.2\, \mathrm{MeV}\,,~~~\la 0|\bar{u} u|0\ra =-(225\,\mathrm{MeV})^3
\\
&& \hat{m}=5.5\, \mathrm{MeV}\,,~~~\la 0|\bar{u} u|0\ra=-(245\,\mathrm{MeV})^3
\eea
obtained respectively in Refs. \cite{Leutwyler2} and \cite{Smilga}.
Let us note here that using the formula (34), we may write the condensate
(32) in nuclear matter as
\be
\la \bar u u\ra = \la 0| \bar u u|0\ra\left( 1-\frac{\sg\,\on}{m_\pi^2\F^2}\right)\,,
\ee
which vanishes at $\on =2.8\on_0$.

Next, for $\Th^f$ we can write the nucleon matrix element as
\be
\la p|\Th_{\mu\nu}^f|p\ra = 2A^f(p_\mu p_\nu - g_{\mu\nu}m^2/4)\,,
\ee
where the coefficient $A^f$ is determined by the first moment sum rule for
the quark distribution function in deep inelastic scattering \cite{PS}. Evaluated 
at the momentum scale of $1\, \mathrm{GeV}$, it has the value $A^f= 0.62$ 
\cite{Martin}. Then noting the normalization condition, 
$\la 0|\Th^f_{\mu\nu}|0\ra = 0$, we get from Eqs. (31) and (38),
\be
\la \Th^f\ra=\frac{3}{4} m A^f \on
\ee

Finally for the operator $G^2$, we use the trace anomaly to relate 
it to the trace of the full energy momentum tensor $\Th_{\mu\nu}$,
\be
\Th ^\mu_\mu = -\frac{9}{8}G^2 +2\hat{m} \bar{u}u +c\cdot\mathrm{1}
\ee
where we add the $c$-number term to fix again its vacuum normalization, 
$\la 0|\Th_{\mu\nu} |0\ra = 0$. Taking the vacuum and the ensemble expectation
values, we get
\be
\la G^2\ra=\la 0|G^2|0\ra -\frac{8}{9} (m-\sg)\on
\ee  

With the above results, we can subtract out the vacuum contributions from
Eq.(29) for $\Pi (E,0)$ and write the result for the amplitude combination (22) as
\bea
&&\wpi (Q^2)\stackrel{OPE}{\longrightarrow}\nonumber\\
&&\left[-\frac{A}{8\pi^2}Q^2\ln\left(\frac{Q^2}{\mu^2}\right)
+\frac{B m_2}{8\pi^2}\ln\left(\frac{Q^2}{\mu^2}\right)-\frac{2Cm_2}{3Q^2}\right]
\overline{n}\nonumber\\
\eea
where $A,\,B$ and $C$ stand for the constants,
\bea
&&A=\frac{\sg}{\hat{m}}+12\nonumber\\
&&B=5m A^f-\frac{2}{9}(m-\sg)\nonumber\\
&&C=\la0|\bar{u}u|0\ra\left(f\frac{\sg}{\hat{m}}+3\right)
\eea

\section{Sum rule}

It is now simple to take the Borel transform of the spectral and the operator sides 
and get the desired sum rule
\bea
\lm^{*2}&&=\lm^2 e^{m_1^2/M^2}\left[\frac{m-m_2}{2m^*}e^{-m^2/M^2}-
\frac{3}{64\pi^2\F^2m^*}\times\right.\nonumber\\
&&\int_C\frac{dE}{E}f(E)\left\{1-2g_A
\left(\frac{E+m}{E-m}\right)\right\}e^{-E^2/M^2}\nonumber\\
&&-\left.\frac{M^2}{2\lm^2m^*}\left(\frac{M^2}{8\pi^2}AV_2
+\frac{m_2}{8\pi^2}BV_1 +\frac{2Cm_2}{3M^2}\right)\overline n\right]
\eea
where $f(E)$ is given by Eq.~(26) and
\[V_1=1-e^{-W^2/M^2}, ~~ V_2=1-(1+W^2/M^2)e^{-W^2/M^2}\]
The deviation of $V_{1,2}$ from unity represents the contribution from the
high energy region on the spectral side, obtained by continuing the result
for operator expansion to the time-like region. Here $W$ is a parameter 
determining the onset of this continuum contribution. We take 
$W=2\,\mathrm{GeV}$, as assumed for the vacuum sum rules \cite{Ioffe1}. 

Let us recall here that a special feature of the present $QCD$ sum rule
is its sensitivity to the medium dependent quantities, requiring a more
careful saturation of its spectral and operator sides. Since we consider
nuclear matter at zero temperature, it consists only of nucleons. Then, on
the spectral side, the two-particle density dependent contributions can arise from
intermediate states consisting only of $N$ with $\pi, \rho, \omega, \phi, \cdots$. 
As seen from the inequalities (15,16), the range of energy over which these states 
may contribute is highly restricted. Thus at normal nuclear density, these ranges 
are only $1080\, \mathrm{MeV}\leq E\leq 1290\, \mathrm{MeV}$ on the unitary cut and 
$670\, \mathrm{MeV} \leq E \leq 800\, \mathrm{MeV}$ on the Landau cut. Inclusion of 
more particles in the intermediate state will, of course, increase these ranges, but
their contributions will fall off exponentially in the integrand. So 
we include only the continuum contribution from the $\pi N$ intermediate state, 
in addition to the modified nucleon pole. Although the continuum contributions 
over the Landau and the unitary cuts are individually rather big, the two largely 
cancel out in the difference (see Eq.~(18)), leaving the nucleon pole contribution 
to dominate the spectral side. 

The operator side is also well saturated with the
usual low dimension operators. The leading operators, $\overline q q$ and 
$q^\dagger q$ bring in a large contribution, as represented by the term with 
$A$ in Eqs.~(42, 44), compared to those of dimensions four and six retained in 
the sum rule. Also the numerous operators of dimension five and higher do not 
appear to contribute significantly, at least individually \cite{Jin}.

\section {Evaluation}

In presenting numerical results, we observe that among all the parameters,
it is the $\lm$, which enters most sensitively in the sum rule and also suffers 
from the largest uncertainty in its value (Eq. (5)). Under these circumstances, 
we fix $\lm$ by requiring maximal stability of the results against variation 
with respect to the Borel mass in a range similar to that required for the vacuum 
sum rules \cite{Ioffe1}. Numerical 
evaluation shows that bigger the value of $\lm$, the more stable is the result. We 
show this stability in Fig.~3 for different relative densities, taking
$\lm^2=.0012 \, \mathrm{GeV^6}$, the central value and $\lm^2=.0018 \,
\mathrm{GeV^6}$, the largest in the allowed range. (Here we take the
self-energies from Ref. \cite{Haar}; those from Ref. \cite{Brockmann} give
even better stability of the results.) With the latter value of $\lm$,
we see that there is a reasonable plateau up to about normal
nuclear density, vindicating our saturation scheme. At higher densities,
however, the plateau seems to disappear, indicating insufficient saturation.

\bfg
\includegraphics[scale=0.5]{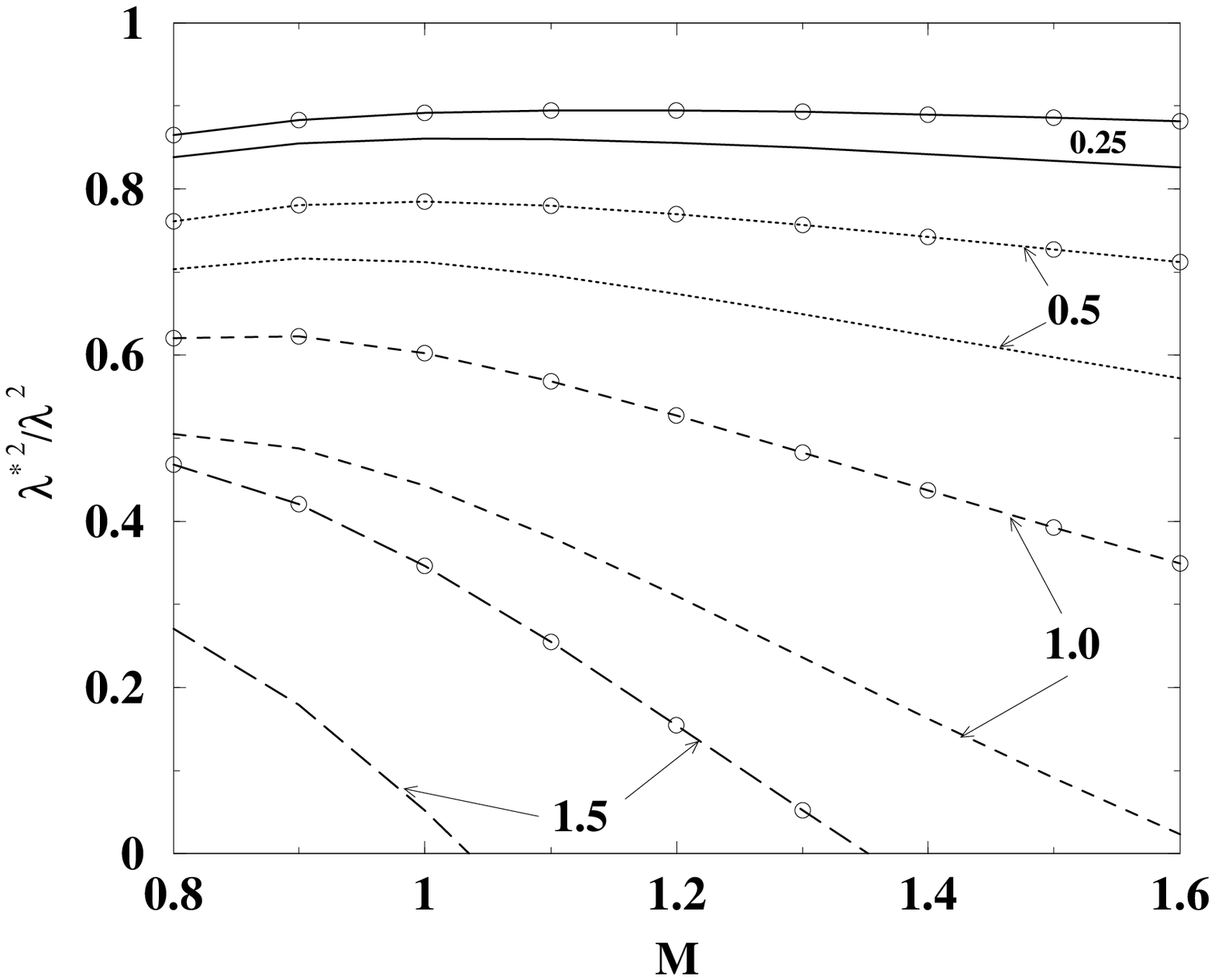}
\caption{Borel mass dependence of $\lm^{*2}/\lm^2$ for four relative  
densities $(\on/\on_0)$, namely $0.25, 0.5, 1.0$ and $1.5$. The curves
spotted without and with circles correspond to $\lm^2 = .0012 \,
\mathrm{GeV}^6$ and $\lm^2 = .0018 \, \mathrm{GeV}^6$ respectively.}
\efg

We next consider the uncertainties in the results from the remaining inputs.
First consider the self-energies. As already stated, these have been
determined by two groups \cite{Brockmann,Haar}, using entirely different
methods based on phenomenological $N\!N$ potentials. Fig.~4 depicts results
using these two sets of self-energies. Clearly it is not a source of any
significant uncertainty in the results, at least up to normal nuclear density.

\bfg[t]
\includegraphics[scale=0.5]{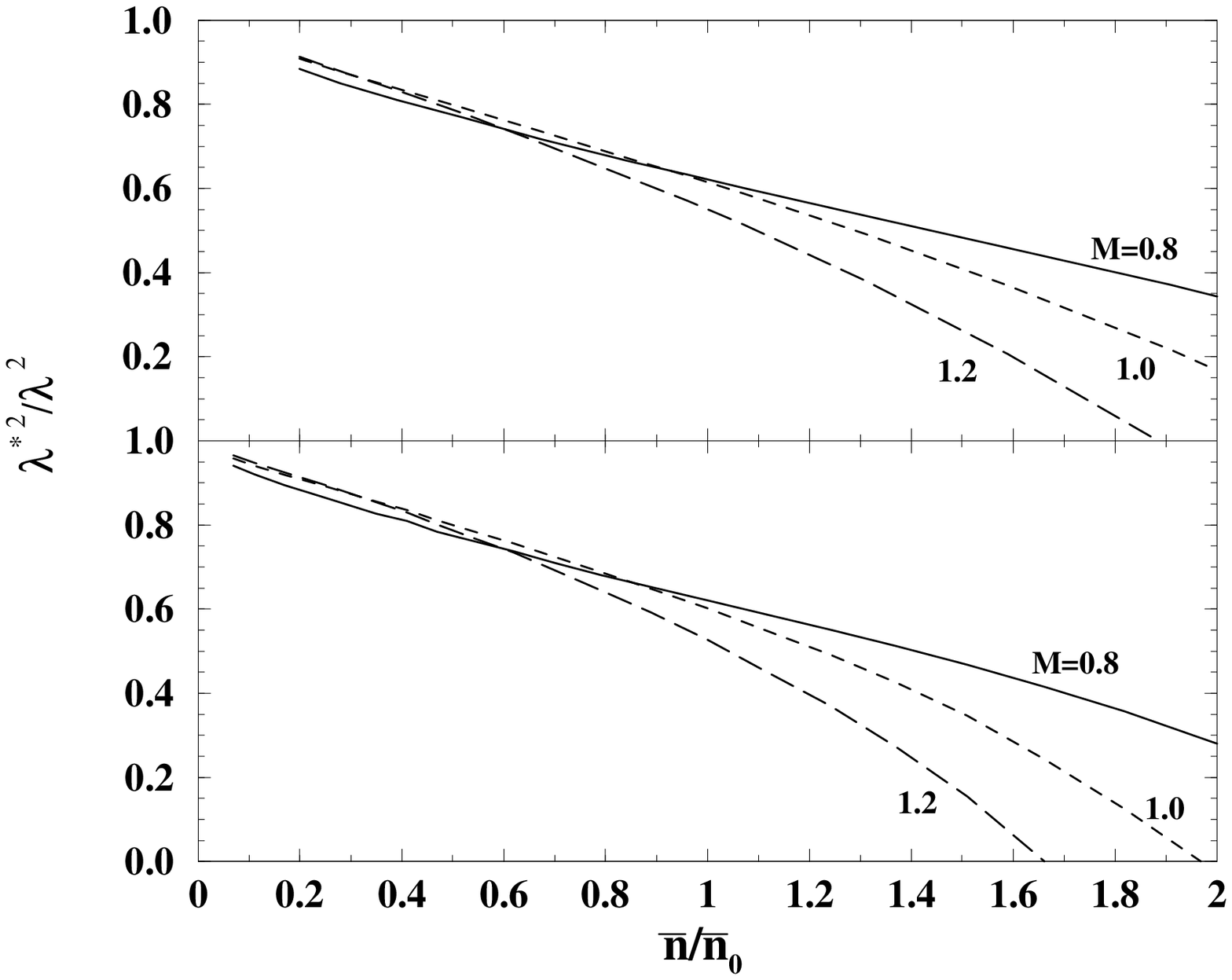}
\caption{Plot of  $\lm^{*2}/\lm^2$ as a function of $\overline n/\overline
n_0$ for three Borel masses $(M)$, namely 0.8, 1.0 and 1.2 GeV. The upper and 
the lower panel show the results using self-energies given in Refs.
\cite{Brockmann} and \cite{Haar} respectively. The self-energies are
available at discrete values of densities, the minimum of which marks the
beginning of our curves here and in Fig. 5 below.}
\efg

Finally we vary the values for the sigma term and for the pair, the quark
mass and the quark condensate. As seen from Fig.~5, the uncertainty in these
parameters again does not give rise to any significant spread in the values
of $\lm^*$. Also the term with the parameter $f$ in Eq.~(43) arising from the
approximation to the four-quark condensate is relatively too small to change
the results appreciably.
We thus show unambiguously a decreasing trend for $\lm^*$ with the rise of
density at least up to normal nuclear density.

\bfg[t]
\includegraphics[scale=0.5]{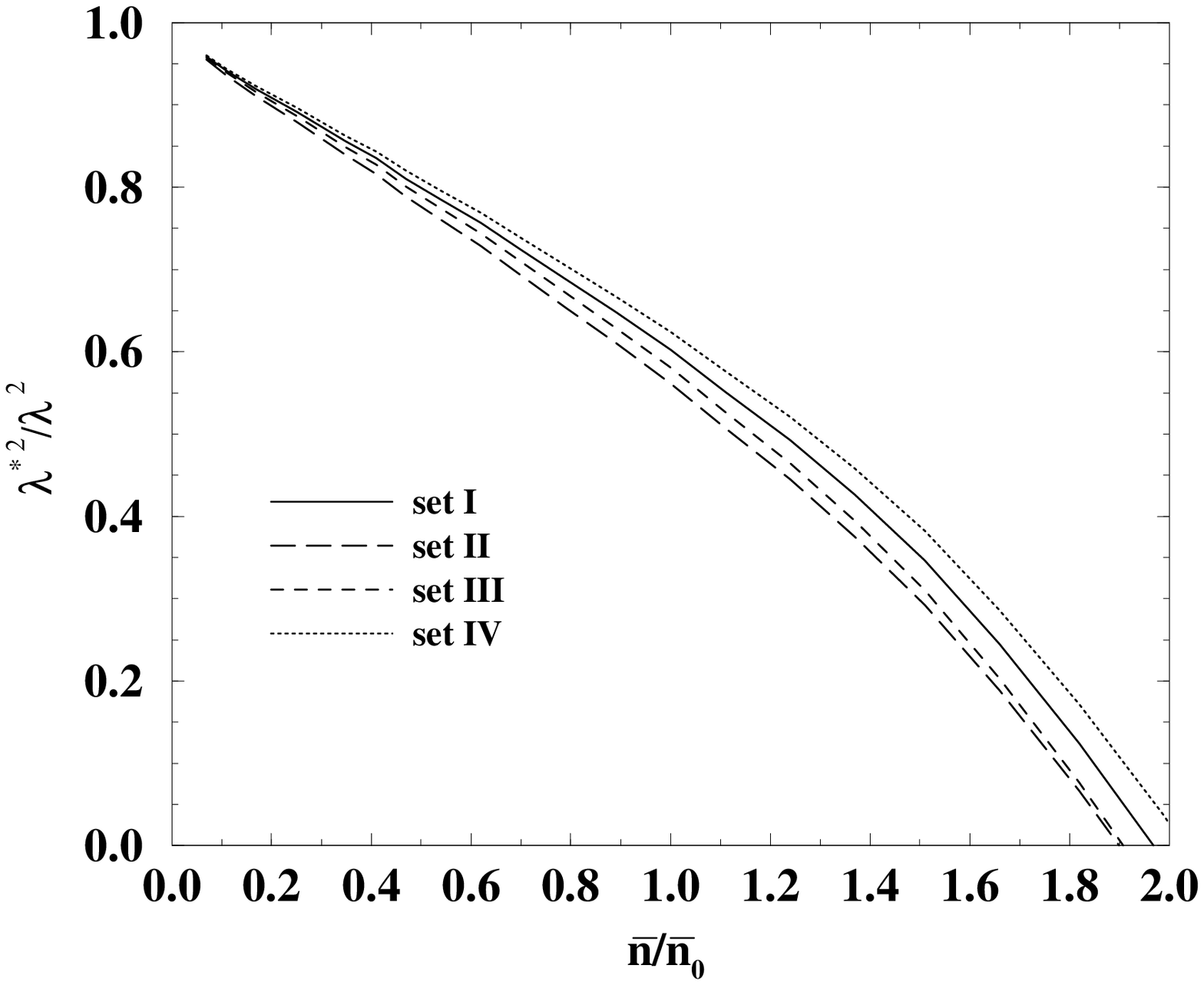}
\caption{Plot of  $\lm^{*2}/\lm^2$ as a function of $\overline n/\overline
n_0$ with four different parameter sets for $\sg$ and the pair, $\hat{m}$ and
$\la  0|\bar{u} u|0\ra$.  I: $\sg=45\,\mathrm{MeV}$, Eq.(35); II:
$\sg=45\,\mathrm{MeV}$, Eq.(36):  III: $\sg=53\,\mathrm{MeV}$, Eq.(35): IV:
$\sg=37\,\mathrm{MeV}$, Eq.(35). We take the Borel mass of $1$ GeV and the 
self-energies from Ref. \cite{Haar}.}
\efg

\section{Discussion}

We describe here a method to find $\lm^*$, the parameter coupling the
nucleon current to the nucleon state in nuclear matter. The method consists of 
writing down a $QCD$ sum rule in this medium for an appropriate combination of 
amplitudes representing the ensemble averaged two-point function of nucleon 
current. With the present scheme of saturating the sum rule, the results are 
reliable up to normal nuclear density, within which one finds definitely a 
decrease of $\lm^*$ with increase in density. For quantitative results beyond 
this density, we have to improve upon our saturation scheme. A test of this 
improvement at a higher density is offered by observing a plateau in $\lm^*$ at 
such density as a function of the Borel mass. If, however, we continue the
present calculation beyond its range of validity, it goes to zero at about 
twice this density.

In looking for additional sources of significant contributions at higher   
densities, we note that as the chemical potential increases, the Landau and
the unitary cuts extend over wider intervals in $E$, so that contributions
from intermediate states with higher thresholds need be included. Also   
if the many operators of higher dimensions contribute with the same sign,
they together may bring in a significant contribution.

It is appropriate here to compare our work with that of Ref. \cite{Cohen},
where similar sum rules are discussed. These authors work for $\vp \neq 0$, 
getting three independent
amplitudes  and hence three sum rules to determine the two self-energies
and the nucleon residue. They, however, find the latter to be poorly
determined by the sum rules. As we point out in footnote 1, there is only
one combination of these amplitudes in the neighbourhood of $\vp =0$, which
has a large nucleon pole term. We thus write a single sum rule for just this
amplitude, supplying the (unknown) self-energies from independent theoretical
determinations. It is this focussing on the single large amplitude that makes
our result for the residue quantitative.

In our present calculation we face a problem in that the coupling parameter
$\lm$ in vacuum itself is not well determined. This, however, is not the
case with the couplings of vector and axial-vector currents with the
relevant particles, though the problem of determining the self-energies of
these particles in nuclear medium still remains. Once the latter problem is
treated properly, calculation of such parameters may well be quantitative. 

The present work is restricted to nuclear matter at zero temperature, so that the
pions appearing in the loop graphs of Figs. 1 and 2 are virtual. It may, however, 
be easily extended to finite temperature, when there will appear real pions,
exciting the nucleons into nucleonic resonances, such as $\Delta (1237)$ with
high degeneracy factors. Then contributions from intermediate states like 
$\pi \Delta$ need also be included in the spectral side of the sum rule. 

As we work with nuclear matter, we ignore electromagnetic interaction.
However, in the realistic case of nuclear density created, for example, in heavy 
ion collisions, it may not be negligible \cite{Ravndal}.

Finally we put together some similar, known results for
current-particle couplings and the quark condensate. Consider first the
pionic medium at low temperature. The coupling parameter $\F$ of the
axial-vector current with pion in vacuum changes to $\F^T$
\cite{Gasser3,Toublan},
\be
\F^T =\F \left( 1-\frac{T^2}{12\F^2}\right)\,.
\ee
The coupling of the baryonic current with nucleon, that we are considering
here, also changes from the vacuum value $\lm$ to $\lm^T$ \cite{Leutwyler1},
\be
\lm^T =\lm\left\{ 1-\frac{(g_A^2+1)}{32}\frac{T^2}{\F^2}\right\}\,,
\ee 
For the quark condensate, we have
\be
\la \bar{q} q\ra^T=\la 0|\bar{q} q|0\ra \left( 1-\frac{T^2}{8\F^2}\right),
\ee
where we keep only the leading term, though it has been calculated up to
$O(T^6)$ \cite{Becher}.

Considering nuclear matter (at zero temperature), the Lorentz invariance
already breaks at leading order for the axial-vector current coupling to
pion,
\be
k_\mu \F \rw k_0 F_\pi^t \de_{\mu 0} +k_i \F^s \de{\mu i}
\ee
giving rise to two decay parameters. They change with nuclear density as
\cite{Meissner,MS2}
\bea
&& \F^t =\F \left\{ 1-(0.26\pm 0.04) \frac{\on}{\on_0} \right\}\\
&& \F^s =\F \left( 1-(1.23\pm 0.07) \frac{\on}{\on_0} \right\}
\eea
Observe that the two parameters have quite different density dependence. But
one can argue \cite{Meissner} that it is the temporal component that
reflects the spontaneous breaking of chiral symmetry. To these pion decay
parameters, we add the result of the present work, 
\be
\lm^*=\lm\left\{1- (0.20\pm 0.04) \frac{\on}{\on_0}\right\}\,,
\ee
obtained as a linear fit to the curves of Fig.~5 up to normal nuclear
density. Also the quark condensate in this medium is given already by
Eq.(37).

All these results are low-density expansions (in pionic and nuclear
media) for the current-particle couplings and the quark condensate. What
is worth noting is that at such densities all the couplings along
with the quark condensate definitely decrease with rise of density. The
significance of this decrease is clear. In vacuum, $\F\, (\lm)$ measures the
overlap of the pion (nucleon) state with those obtained by applying the
axial (nucleonic) charge on the vacuum. In medium their decrease indicates
that the external sources, $A_\mu (x)$ and $\eta (x)$ become gradually weak
in exciting pion and nucleon in the respective media. The decrease of the
quark condensate implies a trend towards restoration of chiral symmetry.

It is interesting to recall here that although $QCD$ is 
believed to be the theory of strong interaction amongst quarks and gluons,
it is asymptotically free at short distances. This feature leads to the common
belief that with the rise of temperature and/or density, hadronic matter will
undergo one or more phase transitions, restoring the spontaneously broken
chiral symmetry and liberating colour to form quark-gluon plasma. The
corresponding order parameters are the quark condensate and the Polyakov
loop, which must be evaluated on the lattice to investigate such
transitions. Any density expansion of the parameters, even if carried to
arbitrarily high order, would not be valid close to phase transition 
\cite{Leutwyler3}.

It is, however, the case that often the above first order formulae give critical
values in qualitative agreement with lattice and exact model calculations.
The best example is $\la \bar{q} q\ra^T$, for which the value given by 
Eq.~(47) is of the same order as the one from the lattice~\cite{Bazavov}.  
Thus at finite temperature, the same value approximately for the coefficients 
of $T^2$ in $\F^T,\, \lm^T$ and
$\la \bar{q} q\ra^T$ tends to support
the expectation that they all go to zero at the same (critical) temperature
\cite{Leutwyler1}. In the same way, at finite nucleon chemical potential, 
we find that the coefficients of $\on$ are again approximately the same for 
all the three quantities, allowing us to expect that they all disappear
together at the same critical density, which is several times away from the 
normal nuclear density. Our speculation has an added importance in that 
quantitative calculation on the lattice proves difficult at finite chemical 
potential \cite{Gavai}.

\section*{Acknowledgement}

One of us (S.M.) wishes to thank Professors H. Leutwyler and F. Ravndal for
discussions. He also acknowledges support from Department of Science and
Technology, Government of India.

\end{document}